\pdfoutput=1

\documentclass[aps,prb,reprint,showpacs,superscriptaddress]{revtex4-1}
\usepackage{graphicx}
\usepackage{graphics}
\usepackage{amsmath}
\usepackage{amssymb}
\usepackage{amsfonts}
\usepackage{dcolumn}
\usepackage{dsfont}
\usepackage{latexsym}
\usepackage{rotating}
\usepackage{color}
\usepackage{latexsym}
\usepackage{bbm}
\usepackage{subfigure}
\usepackage{float}
\usepackage{epsfig}
\usepackage{epsf}
\usepackage{psfrag}
\usepackage{bm}
\usepackage{amsthm}
\usepackage{eucal}
\usepackage{mathrsfs}
\usepackage{url}
\usepackage{braket}
\usepackage{epstopdf}
\usepackage{color} % use colored text in latex

\begin{document}
\title{Effects of Sr-doping on the electronic and spin-state properties of infinite-layer nickelates}
  
\author{Jyoti Krishna}
\affiliation{Department of Physics, Arizona State University, Tempe, AZ 85287, USA}
\author {Harrison LaBollita}
\affiliation{Department of Physics, Arizona State University, Tempe, AZ 85287, USA}
\author {Adolfo O. Fumega}
\affiliation{Departamento de Fisica Aplicada, Universidade de Santiago de Compostela, Santiago de Compostela E-15782, Spain}
\affiliation{Instituto de Investigacions Tecnoloxicas,Universidade de Santiago de Compostela, Santiago de Compostela E-15782, Spain }
\author {Victor Pardo}
\affiliation{Departamento de Fisica Aplicada, Universidade de Santiago de Compostela, Santiago de Compostela E-15782, Spain}
\affiliation{Instituto de Investigacions Tecnoloxicas,Universidade de Santiago de Compostela, Santiago de Compostela E-15782, Spain }
\author{Antia S. Botana}
\email{antia.botana@asu.edu}
\affiliation{Department of Physics, Arizona State University, Tempe, AZ 85287, USA}

\date{\today}
  
\begin{abstract}

The recent discovery of high-T$_{c}$ superconductivity (HTS) in Sr-doped NdNiO$_2$ has sparked a renewed interest in investigating nickelates as cuprate counterparts. Parent cuprates [Cu$^{2+}$: d$^9$] are antiferromagnetic charge transfer insulators with the involvement of a single d$_{x^2-y^2}$ band around the Fermi level and strong $p-d$ hybridization. In contrast, isoelectronic NdNiO$_2$ [Ni$^+$: d$^9$] is metallic with a d$_{x^2-y^2}$ band self-doped by Nd-d states. Using first principles calculations, we study the effect of Sr-doping in the electronic and magnetic properties of infinite-layer nickelates as well as  the nature of the holes. We find that hole doping tends to make the material more cuprate-like as it minimizes the self-doping effect, it enhances the $p-d$ hybridization, and it produces low-spin (S=0, non-magnetic) Ni$^{2+}$ dopants in analogy with the S=0 Zhang-Rice singlets that appear in cuprates. 
\end{abstract}

\maketitle

\section{Background} 
A possible route to address the origin of high-temperature superconductivity (HTS) is to find cuprate analog families, which might
help unveil what is relevant for HTS \cite{norman-RPP}. One plausible strategy to find cuprate analogs is to replace\cite{anisimov} Cu$^{2+}$ with isoelectronic Ni$^{1+}$: d$^9$ . This oxidation state formally takes place in infinite layered nickelates RNiO$_2$ (R= La, Nd) \cite{ikeda, ikeda2, crespin, hayward, hayward_nd}. After 30 years of trying, Sr-doped NdNiO$_2$ has recently been reported to be a superconductor with T$_c$ $\sim$ 15 K \cite{new} and a dome-shaped doping dependence\cite{li2020_dome}. 

As discussed in the literature, the parent phase of 112 nickelates (at d$^9$ filling) is quite different from that of cuprates. Experimentally, transport data indicate that RNiO$_2$ materials are
not insulating and there is no experimental evidence for antiferromagnetic order in
any RNiO$_2$ material \cite{ikeda, ikeda2,crespin, hayward}. In addition, electronic-structure calculations
of RNiO$_2$ indicate significant differences from cuprates
due to the presence of low-lying R-5d states crossing the Fermi level that effectively self-dope the d$_{x^2-y^2}$ band so that the formal electron count for the Ni becomes d$^{8+\delta}$. Also, because Ni is to the left of Cu in the periodic table,  it has an increased charge-transfer energy  \cite{Thomale_PRB2020, prx, pickett, pickett2, Hepting2020} (twice as large than a prototypical cuprate value). However, both in cuprates and infinite-layer nickelates superconductivity appears upon hole doping, so a relevant question to address is how do the above parameters change upon doping as well as the nature of holes.

\begin{figure}
\includegraphics[scale = 0.25]{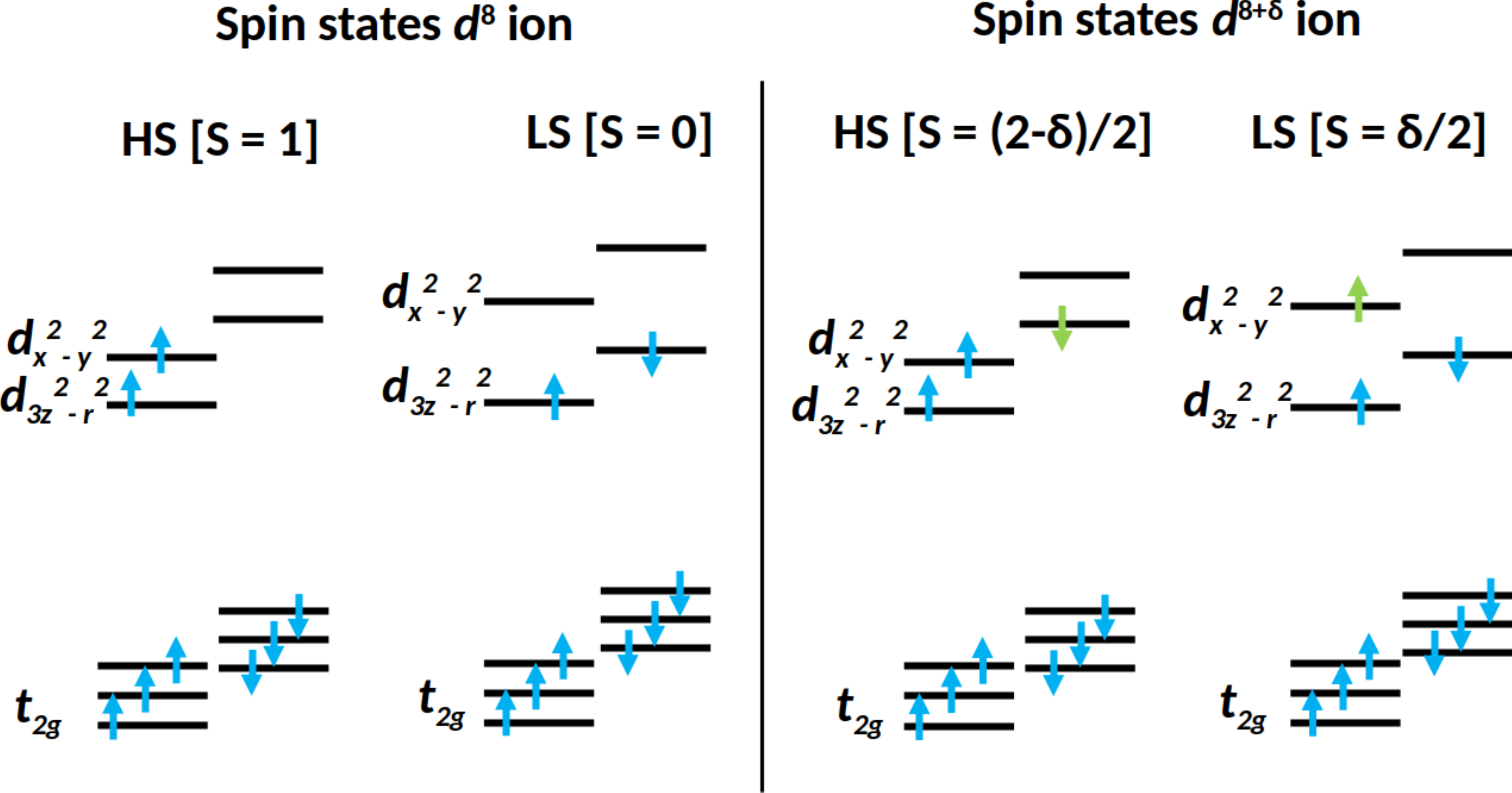}
\caption{Schematic representation of the energy level diagrams for high-spin and low-spin states of Ni$^{2+}$(d$^8$)(left) and Ni$^{2+\delta}$(d$^{8+\delta}$)(right) ion in a square-planar environment. t$_{2g}$ label refers to non-degenerate $xy$, $xz$, $yz$ orbitals for simplicity. Green arrows refer to a partial occupation of that particular orbital.}
\label{fig1}
\end{figure}

In cuprates, one starts with a parent phase in which the Cu$^{2+}$ ions have one active d$_{x^2-y^2}$ orbital hybridized with the $p$ orbitals of the neighboring in-plane oxygens \cite{khomskii_2014}. When such a system is doped by holes, one would formally create states Cu$^{3+}$ (d$^8$) from the initial Cu$^{2+}$ (d$^9$). However, Cu$^{3+}$ is a state with negative charge-transfer gap so the situation is effectively different and holes  predominantly go to the O-p orbitals, that is, the state Cu$^{2+}$(d$^9$)\underline{L} (where \underline{L} stands for a ligand hole, e.g., hole on oxygen). This allows the formation of Zhang-Rice singlets \cite{zhang_rice}. As mentioned above, in nickelates the charge-transfer energy is much larger  \cite{prx} so one may expect Ni$^{2+}$ to be dominant with holes residing on the Ni, rather than in the O-p band.  Ni$^{2+}$ can be in two different spin states, as depicted in Fig. \ref{fig1}: i) A low-spin state (LS) if the electron from the higher-lying d$_{x^2-y^2}$ orbital is removed, in which case the resulting configuration is (t$_{2g})^6$ (d$_{z^2}$)$^2$ with S=0; ii) a high-spin (HS) state if the electron from the d$_{z^2}$ orbital is removed, in which case the resulting configuration is (t$_{2g})^6$ (d$_{z^2})^1$ (d$_{x^2-y^2})^1$ with S = 1 \cite{nat_phys,eff_ham}.  It has recently been argued that hole-doping Nd112 should produce Ni$^{2+}$ with spin S = 1 \cite{lechermann, lechermann2, Petocchi2020NormalSO}. 
Here, we find from first-principles calculations that include explicit Sr-doping in RNiO$_2$ (R= La, Nd) supercells, that a low-spin state (S=0) is preferred instead  due to the large crystal field splitting of the e$_g$ states in a square planar environment. This is consistent with what is found in other square planar layered nickelates \cite{nat_phys,pardo2012,la438_magnetism,la438_prl,la438_CO}. We also find that upon hole doping, the electronic structure of infinite layer nickelates becomes more cuprate-like with reduced charge transfer energy and suppressed self-doping.

\begin{figure}
\includegraphics[scale = 0.55]{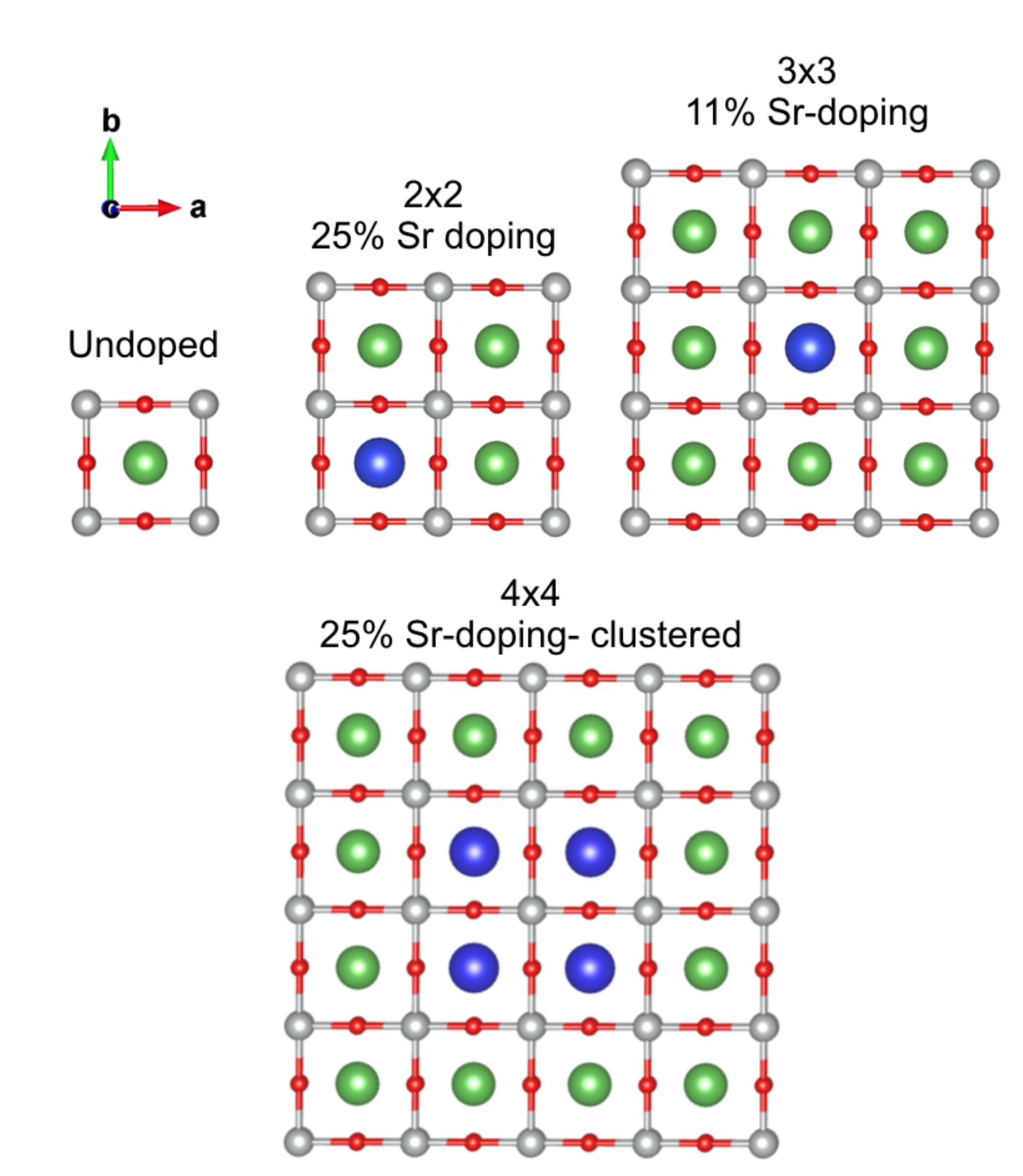}
\caption{Crystal structure of undoped RNiO$_2$ (R=La, Nd)  (a) and Sr-doped RNiO$_{2}$ supercells of different size 2$\times$2 (b), 3$\times$3 (c), and 4$\times$4, corresponding, respectively, to average Ni-d fillings: 8.75, 8.89  and 8.75. R atoms in green, Sr-atoms in blue, oxygen atoms in red, Ni atoms in gray. Note that (d) corresponds to a particular type of 25\% Sr-doping, where all the Sr dopants are clustered around a particular Ni cation, that one can study as the closest possible approximation to a nominally Ni$^{2+}$ impurity.}
\label{fig2}
\end{figure}

\section{Computational Methods} 

 We have performed calculations in both LaNiO$_2$ (with $a$= 3.96 \AA, $c$= 3.37 \AA), and  NdNiO$_2$ ($a$= 3.92 \AA, $c$= 3.28 \AA), two experimentally available members infinite-layer nickelates, in order to check for chemical pressure effects.
Supercells of size 2$\times$2,  3$\times$3 and 4$\times$4 relative to the primitive P4/mmm
cell were employed to study the effect of 11\% and 25\% Sr  doping in both LaNiO$_2$ and NdNiO$_2$.  They give rise to an average Ni-\textit{d} filling of 8.75, 8.89 and 8.75, , respectively (with  4 Sr dopants in the unit cell clustered around a particular Ni atom in the latter, which becomes nominally a Ni$^{2+}$). The corresponding structures are shown in Fig. \ref{fig2}. The structures of the Sr-doped supercells were fully relaxed using the pseudo-potential code Vienna \textit{ab-initio} simulation package (VASP)\cite{vasp1,vasp2} within Perdew-Burke-Ernzerhof version of the generalized
gradient approximation (GGA) \cite{pbe}. Electronic structure calculations were performed using the all-electron, full potential
code WIEN2k \cite{wien2k} based on the augmented plane wave
plus local orbitals (APW + lo) basis set. The missing correlations beyond GGA at Ni sites were taken into account through LDA+U calculations \cite{ldau}. Two LDA+$U$ schemes were used: the `fully localized limit' (FLL) and the `around mean field' (AMF) \cite{sic,amf}. For both schemes, we have studied the evolution of the electronic structure with increasing $U$  ($U_{Ni}$= 1.4 to 6 eV, $J$= 0.8 eV).

\section{structural properties}

The structure of RNiO$_2$ materials consists of NiO$_2$ planes (with 180$^\circ$ O-Ni-O bond angles) that are separated by a single layer of R ions. The Sr-doped structures experience a small increase in volume with respect to the parent materials $\sim$ 1-2 \% for the levels of doping we are considering, up to 25\%. This trend is the expected one as Sr$^{2+}$ is larger than La$^{3+}$ and Nd$^{3+}$. Also, in the structures containing Sr-doping, there is a slight deviation in bond-lengths and bond-angles after structural relaxation.  In particular, there is a distortion of the Ni-O distances in the
NiO$_2$ planes consisting of a modulation of the Ni-O
bond length: shorter around the Ni ions closer to a 2+ state, those proximate to the Sr atoms (1.95-
1.96 \AA) and longer around Ni ions closer to a 1+ state (1.98-1.99 \AA), those away from the dopants. Finally, there is significant buckling close to the Sr-substituted atom with Ni-O-Ni bond angles $\sim$ 175$^\circ$.

\section{electronic structure and magnetism}

\begin{figure}
\includegraphics[width=0.8\columnwidth]{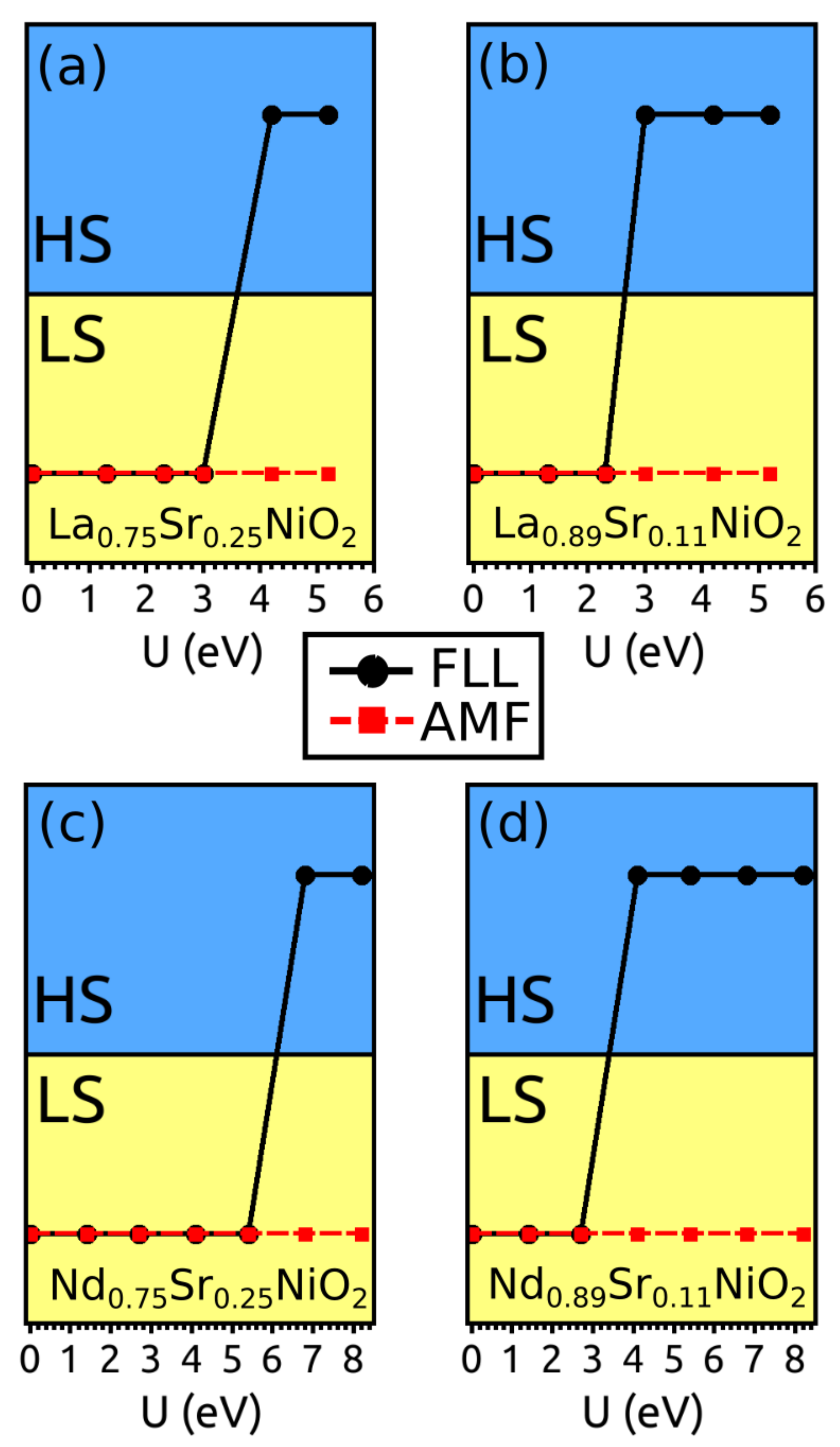}
\caption{Evolution of the lowest-energy spin state (either HS or LS states) with U for the two types of
LDA+U schemes: FLL (fully localized
limit) and AMF (around mean field). Results indicate that only at large U and only for the FLL functional, the HS state could be stabilized, otherwise the LS state is more stable for both dopings studied for the La and Nd cases.}
\label{fig3}
\end{figure}

Using spin-polarized calculations, we have performed a full study of the stability of different spin states for different magnetic configurations in the Sr-doped RNiO$_2$ supercells shown in Fig. \ref{fig2} of size 2$\times$2, 3$\times$3, and 4$\times$4 with effective $d$ fillings of d$^{8.75}$ d$^{8.89}$, and clustered-d$^{8.75}$, respectively. In the smaller cells (2$\times$2, 3$\times$3), two possible spin states can occur in Sr-doped RNiO$_2$.
The (on average) d$^{8+\delta}$ cations sit in a square
planar environment that leads to a large splitting between the d$_{x^2-y^2}$ and d$_{z^2}$ bands.
The crystal field splitting ($\Delta_{cf}$) within the e$_g$ states can be then comparable
to the Hund's rule coupling (J$_H$); if the former is larger, a low spin (LS) state (with S= $\delta$/2 and a moment $\delta$ per nickel) develops
and if the latter is larger, a high spin (HS) state would be more
stable (with S= (2-$\delta$)/2 and a moment 2-$\delta$ per nickel). In contrast, in a large enough cell, a d$^{8}$ configuration can be stabilized for the Ni closer to the Sr dopants. In this situation $\delta$ would effectively be zero  in the above description so that a LS state with S=0 or a HS state with S=1 could be obtained for that specific Ni atom surrounded by first neighbor Sr atoms. This distinction is depicted in Fig. \ref{fig1} with the HS and LS states leading to drastically different properties: the former is non-cuprate like (in that the d$_{z^2}$ states are the dominant ones around the Fermi level) and the latter is more cuprate-like (in that the $d_{x^2-y^2}$ states dominate around the Fermi level). We note that for either LS and HS d$^{8+\delta}$ cases, an AFM coupling between neighboring Ni atoms is always preferred, mediated by the close to half-filling d$_{x^2-y^2}$ band.  

The energetics for these two different spin states in the different supercells have been analyzed within two different LDA+U schemes: around mean field (AMF) and fully localized
limit (FLL). There is a well-known tendency of these different LDA+U flavors to make LS (AMF) or HS (FLL) more stable, respectively\cite{ldau_pickett}. This is a direct
consequence of the double counting term in the AMF scheme giving magnetic states a larger energy
penalty than FLL does. In addition, we have taken into consideration the possible effect of chemical pressure in the stability of HS vs LS state by performing calculations for both La112 and Nd112. We have also studied the evolution with U of the energy differences as it is also expected that a large U would tend to favor the HS state. 
For the 2 $\times$ 2 and 3 $\times$ 3 cells, the magnetic moments are consistent with the above description in terms of Ni-$d^{8+\delta}$ ions with $\mu$=(2-$\delta)\mu_B$ for the HS state and $\mu$=$\delta\mu_B$ for the LS state with $\delta \sim$ 0.8-0.9 (see Table \ref{tablea1} for more details). We note that within FLL, at low U values a HS state cannot be obtained, whereas at high U values (above 4.2 eV), a LS state cannot be converged. Within AMF, at low U values a HS state cannot be obtained either. These trends agree with the tendencies described above for the different LDA+U flavors. Even though in the smaller cells all Ni ions have the same moment (since the number of inequivalent Ni ions is not enough for one of them to nominally achieve a d$^8$ configuration), they do allow us to establish trends for the two different spin states as a function of U.

Figure \ref{fig3} summarizes the phase diagram of the two possible (HS and LS) states as a function of U for AMF and FLL schemes, for both LaNiO$_2$ and NdNiO$_2$ at two different Sr doping levels- 25\% (2$\times$2 supercell), and 11\% ( 3$\times$3 supercell). Within the AMF LDA+U flavor, the LS state is favored, independent of the chosen U value. The FLL flavor also favors the LS state at
low values of U, whereas at high values of U the high-spin solution is favored. These results can again be understood from the fact that the AMF scheme is known to favor the stabilization
of low-spin configurations, whereas FLL tends to favor high-spin configurations, as explained above. These trends are
consistent with earlier results on the trilayer members of the layered nickelate family \cite{nat_phys,pardo2012}. Given that in square planar trilayer nickelates (with the Pr variant also being metallic) the AMF scheme was deemed to provide a better description of the electronic structure in comparison to experiments, the same can likely be anticipated for metallic infinite-layer systems so these results suggest that indeed a LS state would be the most stable situation for these smaller cells. 
The only remarkable difference observed upon a change in R in Fig. \ref{fig3} is that for NdNiO$_2$, the transition to a HS state within the FLL scheme occurs at a slightly larger U when compared to the La material. This result is expected: the Nd-based unit cell is smaller, and the LS state tends to be typically favored at smaller volumes\cite{pardo2012}. 

\begin{figure}
\includegraphics[width=0.85\columnwidth]{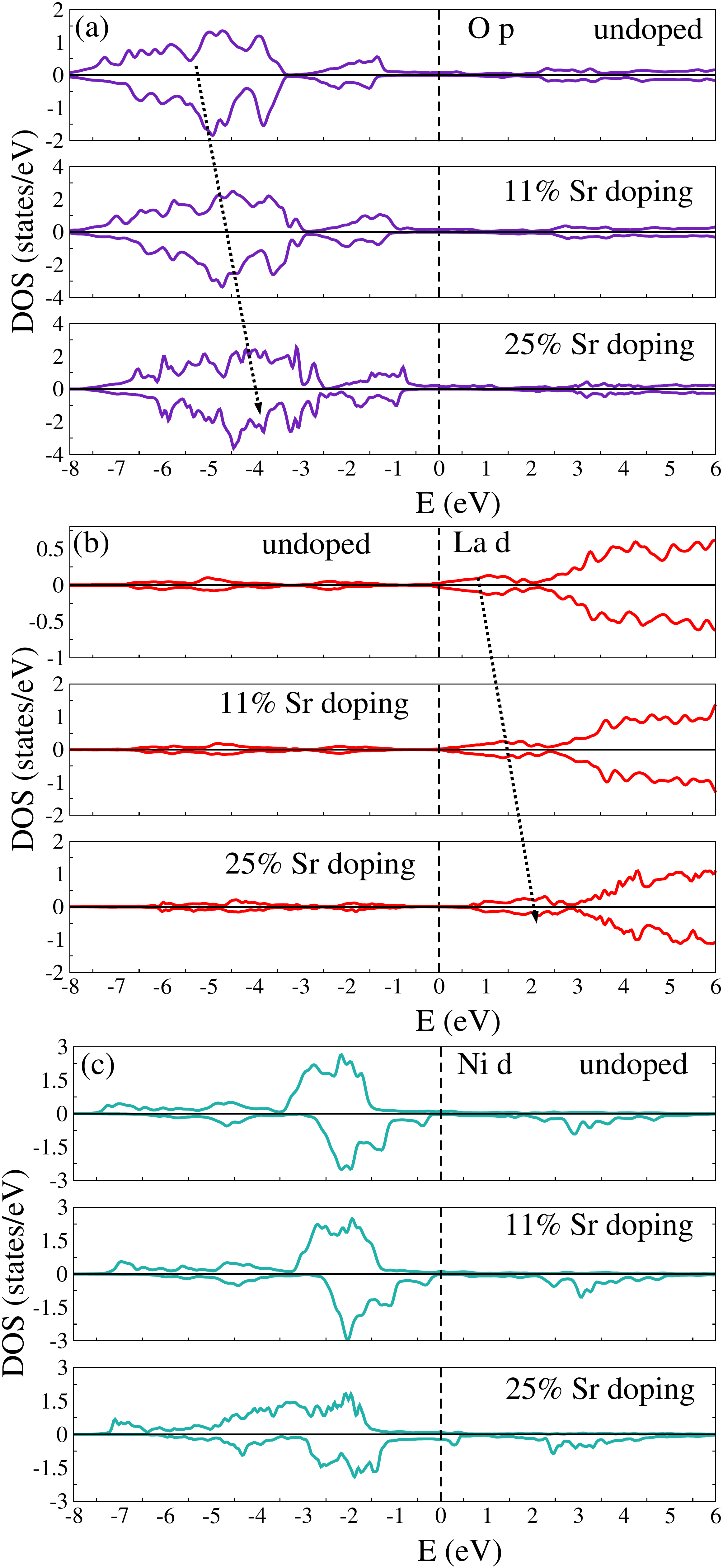}
\caption{Evolution of the orbital resolved DOS for O-p, La-d, and Ni-d states upon increasing Sr-doping in LaNiO$_2$. We see that, as Sr doping is introduced, the centroid of the Ni-d bands does not significantly change while the O-p bands move closer to the Fermi level, effectively reducing the charge-transfer energy. The La d bands move away from the Fermi level, reducing the self-doping effect.}
\label{fig4}
\end{figure}

Using these AMF-LS results for 2$\times$2 and 3$\times$3 Sr-doped cells,  at an effective U= 5 eV we analyze the evolution of the La-d, O-p and Ni-d orbital-resolved densities of states (DOS) upon increasing Sr-doping (see Fig. \ref{fig4}). We show the DOS of LaNiO$_2$ but an identical picture is obtained for its Nd-counterpart.  In a simplified picture, hole-doping would simply shift down the Fermi level, but this is a self-doped system, and hence moving the La-d bands to higher energies would reduce the self-doping effect as a function of Sr doping. We describe how this mechanism works analyzing the actual DOS plots in which the La-d character at E$_F$ can be seen to decrease with doping, as the La-d bands clearly shift to higher energies. This tends to reduce the self-doping effect mentioned above giving a more pure single-band cuprate-like picture. Reinforcing this picture, O-p states are pushed to higher energies with doping. As the centroid of the Ni-d bands does not get shifted noticeably, this reduces the $p-d$ energy splitting (charge-transfer energy) by more than 1 eV when going from the undoped compound to 25\% Sr-doping. We note that the effect on the Ni-d states is more complicated, since on the one hand, Sr introduces holes but on the other, it reduces the self-doping effect from the La-d electron pocket that now moves away from the Fermi level. The overall result is a lower weight of the Ni d$_{z^2}$ band around the Fermi level as Sr-doping is introduced, making the system more cuprate-like in that respect (see Fig. \ref{figa2} for further details). All of the above described trends imply that, as Sr dopants are introduced in RNiO$_2$ materials, some of their electronic-structure features become closer to those of the cuprates: low-spin dopant states, reduced charge-transfer energy, and a single Ni d$_{x^2-y^2}$ band around the Fermi level.

\begin{figure}
\includegraphics[width=\columnwidth]{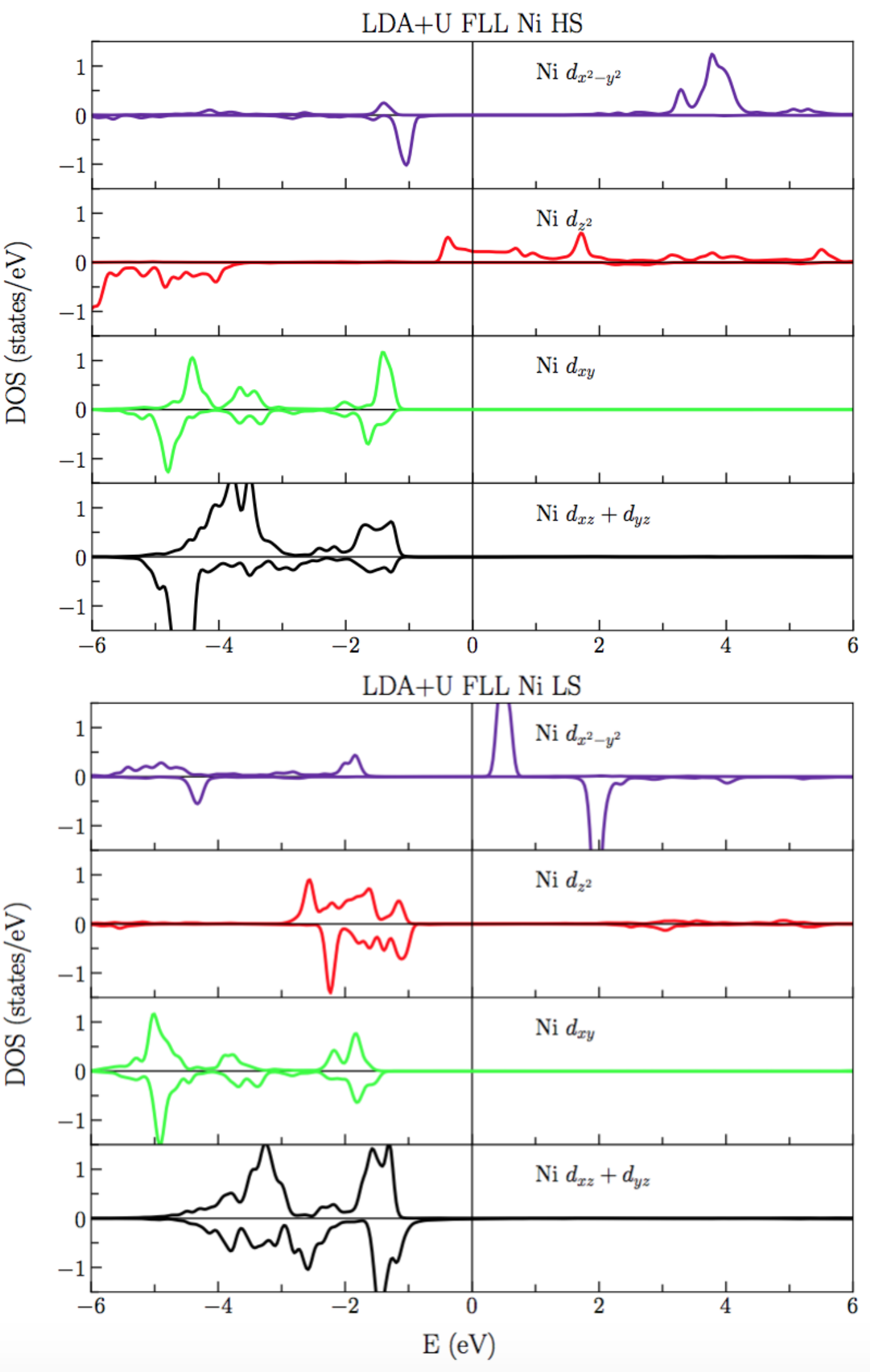}
\caption{DOS of the nominally Ni$^{2+}$ dopant in the 4$\times$4 supercell with 25\% Sr-doping in the two different spin states studied. HS (LS) configuration is shown in the top (bottom) panel. It can be noticed that the lower-energy LS configuration leads to a depletion of Ni d$_{z^2}$ states around the Fermi level.}
\label{fig5}
\end{figure}

We now move to the 4 $\times$ 4 supercells with an average d$^{8.75}$ filling given by a 25\% Sr substitution that allow, via clustering of all the Sr dopants, for one Ni in the cell to be nominally d$^8$ (the Ni atom completely surrounded by first-neighbor Sr cations in Fig.\ref{fig2}). In this scenario, we find that a LS (S=0) state is preferred in both RNiO$_2$ (R= La, Nd), for the Ni ion surrounded by Sr atoms even within FLL- within AMF the low-spin solution is not only the lowest in energy, but a high-spin solution does not even exist, as attempts to start the self-consistency procedure with a S=1 state of Ni$^{2+}$ ion lead to a vanishing magnetic moment. The rest of the Ni atoms preserve the expected magnetic moments (see Table \ref{tablea1} for more details).  Figure \ref {fig5} contrasts the orbital resolved density of states for the nominally d$^8$ Ni cation (that is surrounded by Sr atoms) in both a LS state and a HS state in this 4 $\times$ 4 supercell within FLL. We choose once again to show calculations for Sr-doped LaNiO$_2$ but the situation is identical in the Nd-material. In the low-spin state the t$_{2g}^6$ d$_{z^2}^2$ configuration is clear with the two d$_{x^2-y^2}$ orbitals remaining unoccupied for both spin channels. In the high-spin case, the t$_{2g}$ orbitals are also completely occupied, but now one electron occupies the majority spin d$_{z^2}$ and d$_{x^2-y^2}$ orbitals. The LS state for the Ni$^{2+}$ is strongly connected to a reduction in the d$_{z^2}$ character around the Fermi level. This can be seen in Fig. \ref{fig5} where the Ni d$_{z^2}$ band crossing the Fermi level for the HS state becomes fully occupied, well below the Fermi level for the LS state. This effect is concomitant with the reduction of the La-d self-doping effect upon increasing Sr-doping described above. Overall, the stable LS state solution we find gives then rise to an explicit cuprate-like scenario with planes of S=1/2 ions that are lightly doped with mobile low-spin S=0 ions, a configuration that is directly analogous to the low-spin
S=0 Cu$^{3+}$ ion situation, mediated by O-p holes as explained above.
We note that in other intensively studied nickelates (such as in La$_2$NiO$_4$) the high-spin (S=1) configuration of Ni$^{2+}$ is favored\cite{la2nio4_magnetism_expt,la2nio4_magnetism_rao} and not the low-spin (S=0) configuration as we find here.  La$_2$NiO$_4$ is structurally different to RNiO$_2$ as it preserves apical oxygen atoms and has an octahedral environment for its Ni$^{2+}$ cations. We show here that if the Ni$^{2+}$ ions are forced into a square planar local environment as happens in the 112 materials, they prefer a low-spin d$^8$ (S=0) cuprate-like state instead.

\section{Final remarks}

Using \textit{ab-initio} calculations for Sr-doped RNiO$_2$ (R= La, Nd) systems we have shown that explicit Sr-doping gives rise to important changes in electronic structure and magnetic properties with respect to their undoped counterparts: 1) It reduces the self-doping effect  by shifting the R-d bands up in energy away from the Fermi level leading to a more single-band-like picture, with the Ni d$_{x^2-y^2}$ band dominating. 2) It gives rise to low-spin Ni$^{2+}$ dopants irrespective of the method used, doping level (11\% or 25\%) or rare-earth cation (both La and Nd  yield comparable results). These low-spin dopants make the situation similar to that in cuprates.
3) It drastically reduces the charge-transfer energy  (by up to 1 eV for \%25 doping) bringing it close to values observed in cuprates. Hence, our calculations show that the appearance of superconductivity in infinite-layer nickelates upon Sr-doping might be accompanied by a more cuprate-like electronic structure and spin-states.

\section*{Acknowledgements}
This work is supported by the MINECO of Spain through the project PGC2018-101334-B-C21. A.O.F. thanks MECD for the financial support received through the FPU grant FPU16/02572. AB and JK acknowledge ASU for startup funds.

\clearpage

\appendix

\setcounter{figure}{0}
\renewcommand{\thefigure}{A.\arabic{figure}}
\setcounter{table}{0}
\renewcommand{\thetable}{A.\arabic{table}}

\section{Magnetic moments for all Sr-doped supercells at different U values and orbital resolved Ni-d DOS for different Sr-dopings}

\begin{figure}[!hb]
\includegraphics[width=\columnwidth]{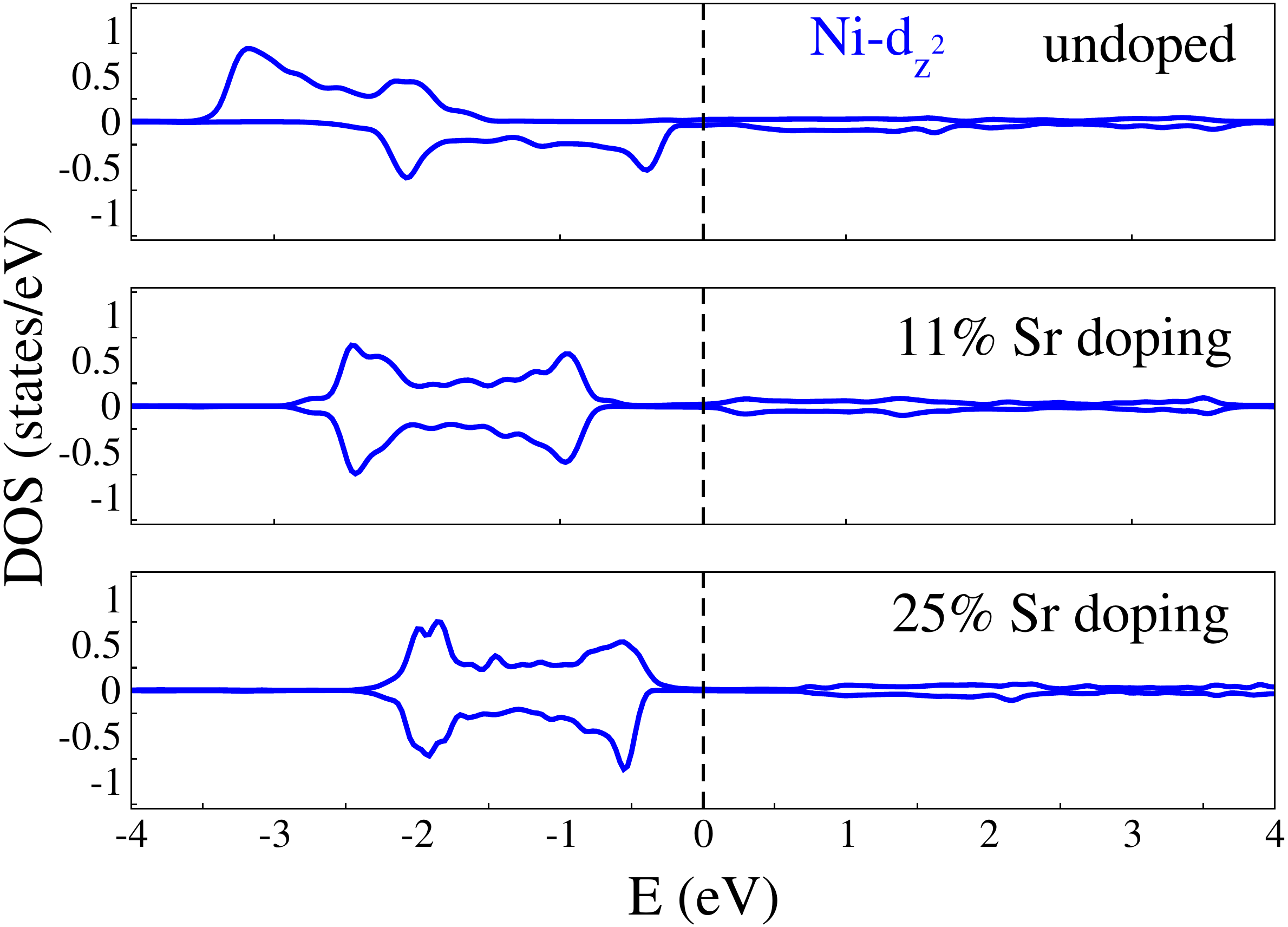}
\includegraphics[width=\columnwidth]{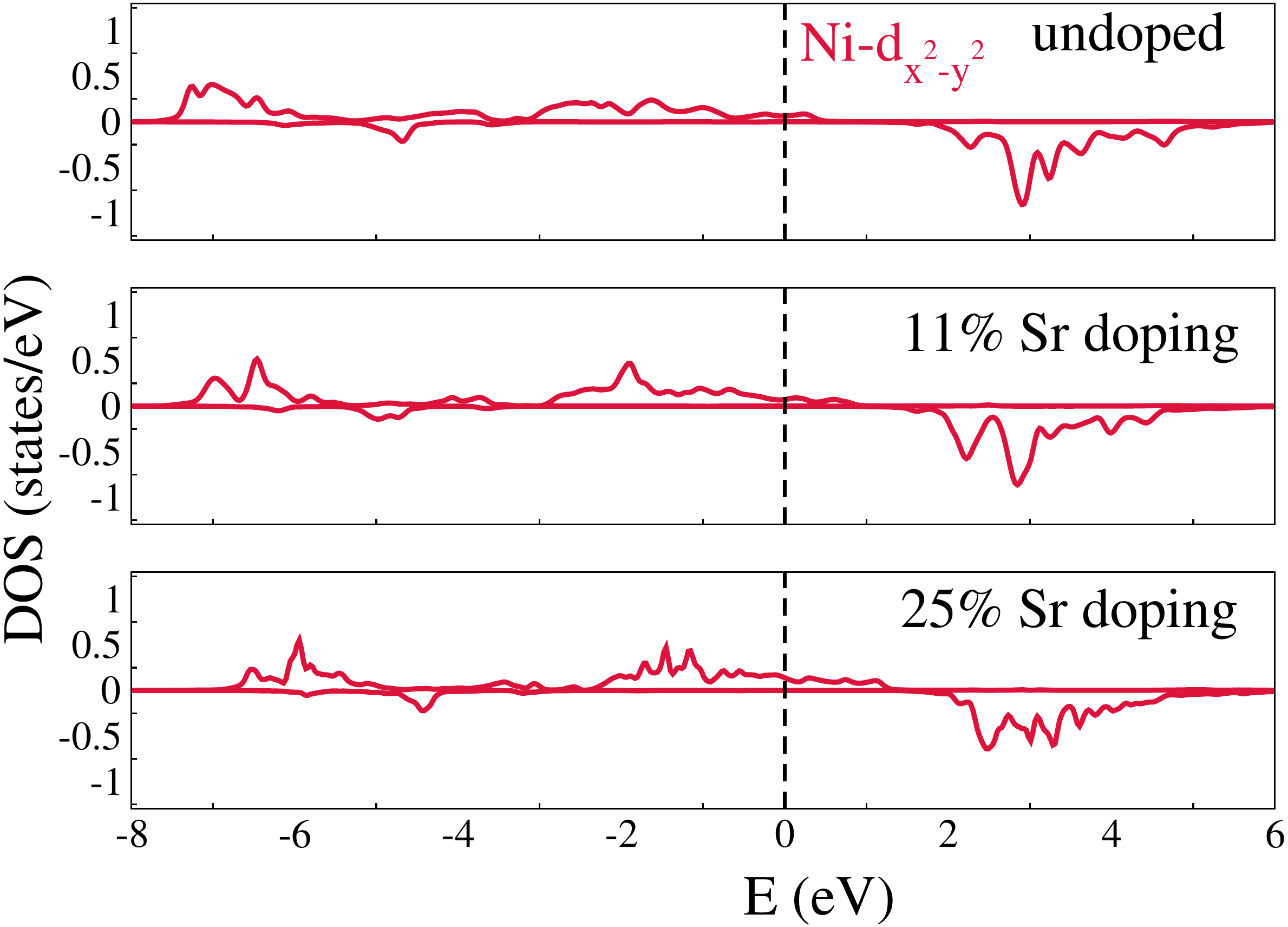}
\caption{Evolution of the orbital resolved DOS for Ni-d$_{z^2}$ and Ni-d$_{x^2-y^2}$ states upon increasing Sr-doping in LaNiO$_2$, obtained using the AMF scheme with U= 5 eV.}
\label{figa2}
\end{figure}

Figure \ref{figa2} shows the orbital-resolved DOS for Ni d$_{x^2-y^2}$ and d$_{z^2}$. As Sr-doping is introduced, the Ni d$_{z^2}$ band moves away the Fermi level, leaving a dominant d$_{x^2-y^2}$ contribution that makes Sr-doped 112 nickelates a more cuprate-like, single-band system than their undoped counterparts.

\newpage

Table \ref{tablea1} shows a summary of the magnetic moments obtained for the different calculations at different Sr-doping concentrations and dopant configurations, for all the U values considered in the two LDA+U schemes utilized, as explained in the main text.

\begin{table}[!hb]
\centering
\caption{Ni atomic magnetic moments (MM) (in $\mu_B$) for the different DFT+U
methods used in this work, for different U values for both HS and LS states. We denote by (-) a solution that cannot be converged.}

\renewcommand{\arraystretch}{1}
\begin{ruledtabular}
\begin{tabular}{lccc}
\multicolumn{1}{l}{\textbf{2 $\times$ 2 La}} \\
\hline
\multicolumn{1}{l}{U-J (eV)} &
\multicolumn{1}{l}{DFT+U flavor } &
\multicolumn{1}{c}{$\mu$ (HS) } &
\multicolumn{1}{c}{$\mu$ (LS) } \\

   \hline
     0 & FLL (AMF) & 0.40 (0.40)  & 0.50 (0.50) \\ 
           1.3 & FLL (AMF) & 1.02 (0.89)   & 0.76 (0.72)\\ 
             2.3 & FLL (AMF) &  1.16 (1.04) & 0.79 (0.72)\\ 
                     3.0 & FLL (AMF) & 1.21 (1.11) & 0.83 (0.72)\\ 
                     4.2 	& FLL (AMF) & 1.25 (1.18) & 1.23 (0.71)\\ 
                     5.2 	& FLL (AMF) & 1.28 (1.25)  & 1.38\\
                     \hline
                     
           \multicolumn{1}{l}{\textbf{2 $\times$ 2 Nd }} \\              
      \hline

     0 & FLL (AMF) & - (-) &  0.29 (0.29)     \\ 
          1.4 & FLL (AMF) & - (-)  & 0.64 (0.60) \\ 
             2.7 & FLL (AMF) &  - (-) & 0.70 (0.62)  \\ 
                    4.1 & FLL (AMF) & - (-) & 0.74 (0.61)  \\ 
         5.4 	& FLL (AMF) & - (-) & 0.81 (0.58)  \\ 
                      6.8 	& FLL (AMF) & 1.42 (-) & - (0.29) \\
         8.2 & FLL (AMF) & 1.53 (-) & - (0.20)\\
                     \hline        
                               \multicolumn{1}{l}{\textbf{3 $\times$ 3 La}} \\ 
                    
\hline
     0 & FLL (AMF) & 0.63 (0.63)  & 0.36 (0.36) \\ 
           1.3 & FLL (AMF) & 0.83 (0.76)    &  0.84 (0.80)\\ 
             2.3 & FLL (AMF) & 0.93 (0.8)  & 0.87 (0.8)\\ 
                     3.3 & FLL (AMF) & 1.15 (0.82)&0.91 (0.79) \\ 
                     4.2 	& FLL (AMF) & 1.24 (1.02) &0.96 (0.77) \\ 
                       5.2 	& FLL (AMF) & 1.30 (1.14)& 1.00 (0.73)\\ 
                     \hline
                     
                               \multicolumn{1}{l}{\textbf{3 $\times$ 3 Nd}} \\ 
                               \hline
     0 & FLL (AMF) & - (-) & 0.52 (0.53)     \\ 
          1.4 & FLL (AMF) & - (-) & 0.42 (0.51) \\ 
             2.7 & FLL (AMF) & - (-) & 0.31 (0.50) \\ 
                    4.1 & FLL (AMF) & 1.41 (-) &  - (0.58)\\ 
         5.4 	& FLL (AMF) & 1.57 (-) & - (0.80) \\ 
                      6.8 	& FLL (AMF) & 1.54 (1.19) & - (0.85) \\
           8.2 	&  FLL (AMF)  &  1.49 (1.51) & - (1.10) \\           
                   \hline
                             \multicolumn{1}{l}{\textbf{4$\times$4 La}} \\ 
                               \hline
              2.3 & FLL (AMF) & 1.10 (-) & 0.74 (0.70) \\ 
             3.3 & FLL (AMF) &1.20 (-) &  0.85 (0.80) \\ 
                     4.3 & FLL (AMF) &1.25 (-)  & 0.95 (0.82)  \\
                     5.2 & FLL (AMF) &1.37 (-) & 1.10 (0.26) \\ 
                     \hline

                             \multicolumn{1}{l}{\textbf{4$\times$4 Nd}} \\ 
                              \hline
                              0 & FLL (AMF) & - (-)  & 0.46 (0.46) \\ 
         1.3 & FLL (AMF) & 1.01 (-)   & 0.75 (0.71)\\ 
           2.3 & FLL (AMF) &  1.14 (1.02) & 0.78 (0.72)\\ 
                   3.0 & FLL (AMF) & 1.19 (1.09) & 0.81 (0.72)\\ 
                    4.2 	& FLL (AMF) & 1.28 (1.16) & 1.20 (0.70)\\
                     5.4 	& FLL (AMF) & 1.57 (1.19) & - (0.79) \\
                     6.8 	& FLL (AMF) & 1.53 (1.19) & - (0.85) \\

\end{tabular}
\end{ruledtabular}
\label{tablea1}
\end{table}

\end{document}